\documentclass[11pt]{article}

\usepackage{graphicx}
\begin{document}
\title{\bf Heterogeneity of the Population of Open Star Clusters in the Galaxy}

\author{{M.\,L.~Gozha, T.\,V.~Borkova, V.\,A.~Marsakov} \\
{Southern Federal University, Rostov-on-Don, Russia}\\
{e-mail: gozhamarina@mail.ru, borkova@ip.rsu.ru, marsakov@ip.rsu.ru }}
\date{accepted \ 2012, Astronomy Letters, Vol. 38, No. 8, pp. 506-518}

\maketitle

\begin{abstract}

Based on published data, we have compiled a catalogue of
fundamental astrophysical parameters for 593 open clusters of the Galaxy.
In particular, the catalogue provides the Galactic orbital elements for 500
clusters, the masses, central concentrations, and ellipticities for 424 clusters,
the metallicities for 264 clusters, and the relative magnesium abundances
for 56 clusters. We describe the sources of initial data and estimate the
errors in the investigated parameters. The selection effects are discussed.
The chemical and kinematical properties of the open clusters and field thin-disk
stars are shown to differ. We provide evidence for the heterogeneity of the
population of open clusters.

\end{abstract}


\maketitle

\section{Introduction}

Open star clusters are traditionally considered to be typical 
representatives of the Galactic thin disk and are used to analyze 
the various aspects pertaining to the structure, chemical composition, 
dynamics, formation, and evolution of this subsystem. It is believed 
that since such parameters as the distances, metallicities, and ages 
are determined for clusters much more accurately than they are for 
single field stars, open clusters trace better the properties of the 
Galactic disk (Piskunov~et\,al.~2006). These stellar systems have a low 
central concentration, are rather weakly gravitationally bound, are 
continuously subjected to a force from massive clouds of interstellar 
gas, and, therefore, are very short-lived systems (Pfalzner~2009). 
Nevertheless, clusters with ages of more than 1 Gyr are also encountered
among them. Unfortunately, the absence of young hot stars, high 
dispersiveness, and often a small number of stars in dynamically highly 
evolved old open clusters produce a significant observational selection 
against them. Such clusters are more difficult to distinguish against 
the background of field stars, and this is one of the reasons why so 
few of them are known. This effect is rather difficult to take into 
account, but it should be taken into consideration in the conclusions 
drawn from the properties of observable open clusters.

In recent years, open clusters have ceased to be regarded as a 
homogeneous population, i.\,e., as having been formed exclusively from 
the interstellar matter genetically related to the previous populations 
of thin-disk stars. For example, Marsakov et al. (1990) showed more 
than two decades ago that the field thin-disk stars and open clusters 
occupy slightly overlapping regions in the age-metallicity diagram, and 
the open clusters at the same age, on average, have a lower metallicity.
As a result, it turned out that low-metallicity ($\textrm{[Fe/H]} < -0.2$)
open clusters are being formed very intensively even at present, while 
virtually no field stars with such a metallicity are born now. As a consequence,
the following conclusion was reached: ''the excess of low-metallicity clusters
most likely results from the fall of a metal-poor gas to the disk and 
the succeeding starburst (for example, due to the interaction with the 
Magellanic Stream)''.

Numerous studies point to the absence of a correlation between the 
properties of open clusters and field thin-disk stars. For example, 
in contrast to the field stars by which a monotonic radial metallicity 
gradient is usually detected (see, e.\,g., Cescutti~et\,al.~2007), Twarog~
et\,al. (1997) showed for open clusters that the metallicity decreases 
with increasing Galactocentric radius with a jump-like transition to 
$\Delta \textrm{[Fe/H]} \approx -0.3$ near 10~kpc. In more recent 
papers (see, e.\,g.,~Friel~et\,al.~2002; Yong et\,al.~2005; 
Magrini~et\,al.~2009), it was shown that there is not a jump but a 
sharp decrease in the slope of the gradient when passing through 
12~kpc, which was explained by satellite accretion onto the outer disk. 
Yong~et\,al.~(2005) investigated the abundances of several chemical 
elements in five old distant open clusters and, in particular, 
concluded that the clusters in the outer Galactic disk could well be 
formed through the stimulation of star formation by a series of 
captures of interstellar matter from dwarf galaxies. Generally, 
estimates show that about 10\,\% of the field thin--disk stars owe 
their origin to open clusters (see, e.\,g., Piskunov~et\,al.~2006). 
(This is without allowance for the numerous "embedded" clusters being 
born with single stars in regions of current star formation and being 
rapidly disrupted.) However, having reconstructed the initial mass 
function of open clusters, Roser~et\,al.~(2010) recently concluded that 
clusters provided the formation of about 40\,\% of the field stars in 
the entire evolution time of the Galaxy; not all of them must 
necessarily turn out to be thin--disk stars. For example, Kroupa~(2002) 
showed how the fast stars leaving the clusters mainly in the period of 
their formation from the parent molecular cloud could thicken the 
Galactic disk and even form a thick disk.

In their comprehensive paper, Wu~et\,al.~(2009) analyzed in detail the 
kinematics and Galactic orbits of 488 open clusters. They found the 
dispersions of all cluster velocity components to increase 
monotonically with age and hypothesized that this resulted from the 
action of dynamical Galactic disk heating effects. They also drew 
attention to the heterogeneity of the population of open clusters: in 
particular, they concluded from their comparison of the orbital 
eccentricities for clusters and thick--disk giants that 3.7\,\% of the 
clusters in their sample belong to the thick disk. In addition, they 
showed that the distribution of clusters in heavy-element abundances is 
bimodal with a dip near $\textrm{[Fe/H]} \approx -0.2$, although they 
pointed out that 
such a distribution could result from metallicity incompleteness of the 
cluster sample. Nevertheless, Vande Putte~et\,al.~(2010) adopted this 
property as evidence for lower--metallicity clusters being isolated by 
assuming an ''unusual'' origin for them. Owing to the appearance of new 
extensive catalogues of highly accurate data, by investigating the 
eccentricities of the Galactic orbits (e) and the maximum distances of 
the orbital points from the Galactic plane ($Z_{max}$) for~481 clusters 
and the metallicities for a hundred and a half clusters, 
Vande Putte~ et\,al.~(2010) revealed about ten objects of an 
extragalactic origin as a result of the interaction of high--velocity 
clouds with the gas disk. They deemed the clusters in which at least 
one of the two investigated orbital elements was abnormally high for 
thin--disk objects and the metallicity was low ($\textrm{[Fe/H]} < -0.2$)
to be such clusters. They deemed the remaining low--metallicity open 
clusters that exhibited no deviations of the orbital elements from the 
mean ones for the bulk of the clusters to have originated from the 
interstellar matter that fell from the outer parts of the Galaxy 
(or was captured from disrupted satellite galaxies). In addition, they 
assumed several clusters with a nearly solar metallicity but with high 
eccentricities or $Z_{max}$ to have been produced by the interaction of 
globular clusters with the disk. The possibility of the formation of 
open clusters with different orbits and metallicities by the listed 
mechanisms triggering star formation was also demonstrated by 
theoretical modeling (Comeron and Torra~1992; Wallin~ et\,al.~1996; 
Martos~et\,al.~1999; Levy~2000).

In their recent paper, Lepine~et\,al.~(2011) explained the abrupt 
decrease in the metallicity of open clusters at a Galactocentric 
distance exceeding the solar one by 1 kpc found by 
Twarog~et\,al.~(1997) by the existence of a corotation ring-shaped 
gap in the gas density, which isolates the inner and outer gaseous 
regions of the Galaxy from each other. (According to their 
calculations, it is at this distance that the rotation velocities of 
the spiral density wave and the Galactic disk coincide.) As a result, 
the inner and outer regions of the Galactic disk chemically evolve 
independently, leading to a break in the radial metallicity gradient 
at this Galactocentric distance. In their opinion, the processes near 
the corotation radius also give rise to clusters with eccentric orbits. 
However, they point out that some of the clusters in the outer disk are 
most likely extragalactic in origin.

These studies are devoted to a comprehensive statistical analysis of 
the relationships between the physical, chemical, and 
spatial--kinematical characteristics of open clusters and nearby field
stars aimed at revealing clusters of different natures as well as 
characteristic parameters and patterns in the various populations of 
Galactic open star clusters. An extensive catalogue of all the 
available parameters of open clusters is primarily needed to 
accomplish the formulated task.


\section {Initial data}

We took the latest version 3.1 (November~2010) of the catalogue 
compiled by Dias~et\,al.~(2002) as the main source of data for open 
clusters. This is currently the most complete catalogue of open star 
clusters that is continuously updated and supplemented with data from 
the most recent publications. The shortcomings of the catalogue include 
the principle of its construction based on the compilation of data by 
many authors that used different criteria, calibrations, techniques, 
and instruments. At the same time, the authors of the catalogue neither 
analyze nor average the data but select the most accurate (from their 
viewpoint) parameter determinations. As a result, the cluster 
parameters in the catalogue turned out to be inhomogeneous. 
However, Paunzen and Netopil~(2006), who used the averaged values in 
their study, showed the results obtained from the data of their 
catalogue and the first version of the catalogue by Dias~et\,al.~(2002) 
to have the same statistical significance. Since our goal is a 
comprehensive statistical analysis of the chemical, physical and 
spatial--kinematical properties of the population of open clusters, we 
deemed it possible to take the data from this most voluminous catalogue 
as a basis while supplementing it with other necessary cluster parameters.

{\bf \em Positions and Velocities}. 

The version of the catalogue by Dias et\,al. (2002) that we use presents
data for 2140 clusters; the distances were determined for 1309 
clusters, while the proper motions and radial velocities are 
simultaneously known for 485 clusters. Since we use the Galactic 
orbital elements calculated by Wu~et\,al.~(2009), Vande Putte~et\,al.~(2010),
and Magrini~et\,al.~(2010) for 500 clusters, we took the distances, 
proper motions, and radial velocities for them from the corresponding 
papers (for more details, see below). Based on the available data, we 
calculated the rectangular coordinates ($x, y, z$) and the space 
velocity components ($V_R, V_\Theta, V_Z$) in cylindrical 
coordinates corrected for the solar motion relative to the local 
standard of rest (LSR), where $V_R$~ is directed to the Galactic 
anticenter, $V_\Theta$~ is in the direction of Galactic rotation, 
and $V_Z$ is directed toward the North Galactic Pole. We took the solar 
motion relative to the LSR to be 
$(U, V, W)_\odot = (11.1, 12.24, 7.25)$~km\,s$^{-1}$  
(Schonrich et\,al. 2010), the solar Galactocentric distance to 
be 8.0\,kpc, and the LSR rotation velocity to be 220~km\,s$^{-1}$. 
The mean relative error of the distances for all clusters was found by 
Wu~et\,al.~(2009) to be $\approx 20\,\%$. The results by Paunzen and 
Netopil (2006) are also consistent with this estimate; they showed the 
error to be less than 20\,\% for 80\,\% of the 395 clusters for which 
three or more independent distance determinations were found. The space 
velocities are generally determined only for clusters in which the 
radial velocities and proper motions for more than five stars can be 
measured. We calculated the mean error of the velocity components using 
data from Vande Putte~et\,al.~(2010) to be $\approx 10.0$~km\,s$^{-1}$ .

{\bf Galactic Orbital Elements}. 

For 488 clusters with all three velocity components, we provided the 
orbital eccentricities (e) and the maximum distances of the orbital 
points from the Galactic plane ($Z_max$) calculated for them by 
Wu~et\,al.~(2009) from these data. We took the same parameters for nine 
more clusters (NGC 433, NGC 1513, Collinder 220, Loden 807, NGC 6249, 
Berkeley 81, IC 1311, NGC 7044, Berkeley  99) from Vande Putte~et\,al.~(2010)
and for three more clusters (NGC 6253, NGC 6404, NGC 6583) from 
Magrini~et\,al.~(2010). (Since Magrini et al. (2010) used a solar 
Galactocentric distance slightly different from that of the other two 
papers in their calculations, 8.5\,kpc against 8.0\,kpc, we reduced the 
radii provided by them by 0.5\,kpc.)

Analysis of the errors in the orbital elements of open clusters given 
in the original papers showed that they depend mainly on the errors in 
the distances to them. Therefore, the largest errors are generally 
obtained for the clusters farthest from the Sun. If the ten distant 
clusters with the largest uncertainties in their orbital elements 
(marked by the asterisks in the table) are excluded from our analysis, 
then the mean error in the apogalactic orbital radii and its dispersion 
for the remaining clusters will be $\epsilon (R_а) = 0.35$ and 
$\sigma(\epsilon)= 0.21$\,kpc. The corresponding values for the 
perigalactic orbital radii are (0.31 and 0.06)\,kpc, for the maximum 
distances from the Galactic plane are (0.09 and 0.04)\,kpc, and for the 
eccentricities are 0.02 and 0.01. (Our analysis showed that the errors 
of all orbital elements for clusters closer (farther) than 1\,kpc from 
the Sun are smaller (larger) than these values, on average, by a factor 
of 1.5 - 2. In contrast, the relative errors of the most informative 
orbital elements, $e$ and $Z_{max}$, can reach 100\,\% for the most 
distant clusters.)

{\bf \em Cluster Ages} 

The cluster ages are not always determined with confidence. In the 
latest version of the catalogue by Dias~et\,al.~(2002), they are given 
for 1269 open clusters. The ages of distant clusters, for which the 
metallicities are unknown (the solar metallicity is assigned to them) 
and the main-sequence turnoff points are not identified, turn out to be 
least accurate. According to the estimates by Kharchenko et\,al.~(2005a),
the errors of the cluster ages on a logarithmic scale turn out to be, 
on average, $\epsilon_{log t}
\approx 0.20 - 0.25$ in this case. Paunzen and Netopil (2006) showed 
that the ages for only 11\,\% of the 395 clusters with three or more 
independent age determinations have errors less than 20\,\%, while the 
errors for 30\,\% of the clusters exceed 50\,\%. They also argue that 
the individual age determinations given in the first version of the 
catalogue by Dias~et\,al.~(2002) also have similar error statistics. In 
contrast, in the latest version of the catalogue that we use, the ages 
were determined slightly more accurately. We see from the above 
estimates that the absolute errors increase with age.

{\bf \em Physical Cluster Parameters} 

We took the physical parameters of the clusters from three catalogues 
of the same team of authors, where these parameters are successively 
estimated by analyzing the spatial distribution of stars in the 
clusters. The mass is the most important characteristic of a cluster 
that determines many aspects of its subsequent existence. We took the 
masses for 424 clusters from the catalogue by Piskunov~et\,al.~(2008), 
where they were estimated from the mean sizes of their semiaxes based 
on King's empirical model. The mean error of the logarithm of the mass 
is about 0.28, which corresponds to a mean relative error of 
$\approx11$\,\%. We calculated the central concentrations $C = lg(r_{cl}/r_{co})$,
where $r_{cl}$ is the angular radius of the cluster and $r_{co}$ is the 
angular radius of its central core, for 424 clusters based on data from 
Kharchenko et\,al.~(2005a, 2005b). These authors use a catalogue of 
2.5 million stars to determine the angular radii and rely on the counts 
of stars brighter than $12^m$. The measurement errors of the 
corresponding radii are not given in their papers, but, according to 
the estimates of the accuracy of the King radii by the same authors 
(see Piskunov~et\,al.~2007, 2008), the relative errors for the 
overwhelming majority of clusters lie within the range from 10\,\% 
to 50\,\%. The ellipticities, the ratio of the difference between the 
semimajor and semiminor axes to the semimajor axis, for 424 clusters 
were taken from the catalogue by Kharchenko~et\,al.~(2009). The mean 
absolute error of the ellipticity in the catalogue turned out to be 
rather large, $\approx 0.21$\,\%, i.\,e., the mean relative error is 
$\approx 60$\,\%.

{\bf \em Metallicities}. 

The largest number of metallicity determinations is given in 
Dias et\,al.~(2002), 179, and Paunzen et\,al.~(2010), 188. (We excluded 
two clusters from the table in Paunzen et al. from consideration, 
because, according to the conclusion reached by Dias et al. (2002), 
Berkeley 42 is a globular cluster, while Ruprecht 46 is an asterism.) 
In version 3.1 of their catalogue, Dias et al. provided the most 
reliable (from their viewpoint) values of [Fe/H] from various sources 
determined both spectroscopically and photometrically. 
Paunzen et\,al.~(2010) simply averaged the photometric metallicity 
determinations of different authors. The [Fe/H] determinations for 
110 clusters are given in both papers and correlate between themselves 
well enough. Assuming the spectroscopic metallicity determinations to 
be more reliable, we included the spectroscopic values from 
Dias et\,al.~(2002) in the first place, the averaged photometric values 
from Paunzen et\,al.~(2010) in the second, and, subsequently, all of 
the values that fell into only one of the lists.
For NGC 188 and Berkeley 39, instead of the metallicities 
(Friel et\,al.~2002) given in the catalogue by Dias~et\,al., we used 
their new values obtained by Friel et\,al.~(2010) based on high--resolution
spectroscopy. We used the values of [Fe/H] for NGC 1193 from the same
paper. Furthermore, we added the metallicity determinations for 
several clusters from Akkaya et\,al.~(2010), Magrini et\,al.~(2010), 
Pancino et\,al.~(2010), and Fossati et\,al.~(2011). In total, our list 
contains 264 clusters with known metallicities. The mean error in the 
metallicity that we determined using the data from these papers is 
$\varepsilon\textrm{[Fe/H]} \approx 0.10$. Our checking showed that 
the distribution of discrepancies between the photometric and 
spectroscopic determinations for the same clusters also exhibits 
exactly the same dispersion.

{\bf Relative Magnesium Abundances} 

We managed to find 81 $[Mg/Fe]$ determinations for 56 clusters in 43 
sources from 1981 to 2011. (We used only the sources in which the 
authors themselves analyzed the accuracies of their determinations in 
different stars and provided the cluster--averaged magnesium 
abundances.) In total, the magnesium abundance was determined in 551 
cluster stars (in one paper for Be 21 and three for NGC 2682, the 
number of stars was not specified). The maximum number of stars 
measured in the cluster is 115 (Hyades), with a mean value of 10 stars 
and a median value of 4 stars. The determinations for NGC 1193 and 
Berkeley 31 were made only from one star. The abundances for ten 
clusters were determined in more than one source, with the maximum 
number of sources being 10 (NGC 2682). For these clusters, we averaged 
the determinations with a weight inversely proportional to the errors 
declared by the authors. The mean error of the relative magnesium 
abundance calculated from the uncertainties of individual 
determinations declared in the sources is 
$\varepsilon\textrm{[Mg/Fe]} \approx 0.07 \pm 0.01$.Comparison of the 
determinations by different authors for clusters with several 
determinations (35 determinations from 29 sources) showed a slightly 
larger dispersion: $\sigma\textrm{[Mg/Fe]} = 0.10 \pm 0.01$. At the 
same time, no systematic shifts between the determinations of different 
teams of authors exceeding this dispersion were found. The ratios 
[Mg/Fe] found and a list of the sources used are given in the table.

For our studies, we compiled a catalogue containing 500 clusters with 
the calculated orbital elements and 264 clusters with the estimated 
metallicities. Since these lists partly overlap, the final catalogue 
contains 593 clusters. The table lists all of the above-described 
parameters found for them  . Column 1 of the table gives the cluster 
name; columns 2 and 3 give the Galactic coordinates (l, b). The 
heliocentric distances $d$, coordinates (x, y, z) in a right-handed 
Cartesian coordinate system, and Galactocentric distances $R_G$ of the 
clusters are listed in columns 4 - 8. The next three columns contain 
the calculated velocity components ($V_R, V_\Theta, V_Z$) in a 
cylindrical coordinate system. The following columns provide the 
Galactic orbital elements of the clusters ($e, Z_{max}, R_a, R_p$). 
The cluster ages are listed in column 16. The physical parameters 
$lg(M/M_{\odot}$), $lg(r_{cl}/r_{co})$, and the ellipticities are 
presented in the succeeding three columns. The values of [Fe/H] and 
references to the source are collected in columns 20 and 21, 
respectively. The relative magnesium abundances [Mg/Fe] and their 
sources are presented in columns 22 and 23. The interpretation of the 
numbers of the references to [Fe/H] and [Mg/Fe] is appended to the 
catalogue. The membership in Galactic subsystems is given in the last 
column of the catalogue. Here, the following notation is used: 
1 -- the thin disk, 2 -- the thick disk, 3 -- the halo.

\begin{table}
\tabcolsep0.5mm
\centering
\caption{%
   Fundamental parameters of open star clusters (a fragment)}
\begin{tabular}{|l|c|c|c|c|c|c|c|c|c|c|c|c|c|c|}
\hline
\multicolumn{1}{|c|}{\bf Name}& 
\multicolumn{1}{|c|}{\bf $l$}&
\multicolumn{1}{|c|}{\bf $b$}&
\multicolumn{1}{|c|}{\bf $d$}&
\multicolumn{1}{|c|}{\bf $x$}&
\multicolumn{1}{|c|}{\bf $y$}&
\multicolumn{1}{|c|}{\bf $z$}&
\multicolumn{1}{|c|}{\bf $R_G$}&
\multicolumn{1}{|c|}{\bf $V_R$}&
\multicolumn{1}{|c|}{\bf $V_\Theta$}&
\multicolumn{1}{|c|}{\bf $V_Z$}&
\multicolumn{1}{|c|}{\bf $e$}&
\multicolumn{1}{|c|}{\bf $Z_{max}$}&
\multicolumn{1}{|c|}{\bf $R_a$}&
\multicolumn{1}{|c|}{\bf $R_p$}
\\
\hline
NGC 6520 &2.8811 &-2.8435&1900&1895& 95&-94&6.11&15.6&213.1&-8.7& 0.07& 0.11&6.3&6.4\\
NGC 6583 &9.2825  &-2.5336&2100&2070&338&-93&5.94&-4.8&215.6&10.1&0.093&0.132&6.56&5.36\\
NGC 6791*&69.9585 &10.9039&5853&1970&5399&1107&8.17&-88.0&217.0&-18.3&0.3&1.16&11.3&6.1\\
NGC 7789 &115.5319&-5.3849&1795&-770&1613&-168&8.92&21.1&158.3&7.8&0.32&0.17&9&4.6\\
NGC 188  &122.8431&22.3841&2047&-1027&1590&780&9.20&-8.2&211.4&-14.7&0.07&0.79&9.3&8.1\\
\hline
\end{tabular}
\end{table}

\begin{table}
\tabcolsep0.5mm
\centering
\begin{tabular}{|l|c|c|c|c|c|c|c|c|c|c|c|c|c|c|}
\hline
\multicolumn{1}{|c|}{\bf $   t$}& 
\multicolumn{1}{|c|}{\bf $log(M/M_\odot)$}&
\multicolumn{1}{|c|}{\bf $log(r_{cl}/r{co})$}&
\multicolumn{1}{|c|}{\bf $Ellipticity$}&
\multicolumn{1}{|c|}{\bf $[Fe/H]$}&
\multicolumn{1}{|c|}{\bf $Ref. [Fe/H]$}&
\multicolumn{1}{|c|}{\bf ${Mg/Fe}$}&
\multicolumn{1}{|c|}{\bf $Ref. [Mg/Fe]$}&
\multicolumn{1}{|c|}{\bf $Subsystem$}&
\multicolumn{1}{|c|}{\bf $name$}
\\
\hline
0.151&1.725&0.34&0.49&-0.25&33&     &         &2&NGC 6520 \\
1.000&     &    &    & 0.37&12&-0.05&       25&1&NGC 6583 \\
4.935&     &    &    & 0.32&12& 0.13& 10,29,34&2&NGC 6791*\\
1.413&3.788&0.43&0.18&-0.24&12& 0.18& 24,30,43&2&NGC 7789 \\
4.285&     &    &    & 0.12&18& 0.26&       24&2&NGC 188  \\
\hline
\end{tabular}
\end{table}

{\bf \em Field Stars} 

In addition to the catalogue of open clusters, we used three more 
lists of field stars. The first is the list containing, in particular, 
the metallicities, Galactic orbital elements, and ages for 2255 
thin--disk stars from Marsakov et\,al.~(2011). It is a sample of nearby 
($<70$~pc from the Sun) F-G stars the probability of whose membership 
in the thin disk is higher than that in the thick one. The sample was 
drawn from the photometric catalogue by Holmberg et\,al.~(2009) based 
on the criteria for selection into the thin disk described by 
Koval'~(2009). This sample is essentially complete for thin-disk 
$F2- G5$ stars within 70 pc of the Sun. Comparison of the photometric 
metallicities from this catalogue with the spectroscopic values of 
[Fe/H] from the catalogue by Borkova and Marsakov (2005) revealed no 
systematic shifts outside errors of $\varepsilon[Fe/H] \approx \pm0.10$ 
(Marsakov et\,al. 2011). About 10\,\% of the stars with errors in the 
ages $>3$~Gyr were removed from the original list; as a result, the 
mean error in the final sample was $\langle t\rangle = \pm1.0$~ Gyr.

The second sample contains 219 nearby thin-disk stars selected from our 
summary catalogue of spectroscopic iron and magnesium abundance 
determinations (Borkova and Marsakov 2005) based on similar criteria. 
Almost all of the magnesium abundances in dwarfs and subgiants of the 
solar neighborhood determined by the method of synthetic modeling of 
high--dispersion spectra that were published before January 2004 were 
collected in the catalogue. The internal accuracies of the catalogued 
metallicities and relative magnesium abundances were 
$\varepsilon\textrm{[Fe/H]} = \pm 0.07$ and 
$\varepsilon\textrm{[Mg/Fe]} = \pm 0.05$, respectively.

The third catalogue is the list containing 135 field Cepheids with the 
distances and spectroscopic iron abundance determinations compiled from 
data from the papers of one group (Andrievsky~et\,al.~2002a - 2002c, 2004;
Luck et\,al. 2003). The typical error of the metallicity in Cepheids 
declared by the authors is $\varepsilon\textrm{[Fe/H]} < \pm 0.1$.


\section{HETEROGENEITY OF THE POPULATION OF OPEN CLUSTERS}

{\bf \em Observational Selection.} 

Recent publications suggest that the currently available 
catalogues of optically visible open clusters are complete 
within 850~pc and, possibly, even 1~kpc (Piskunov et\,al.~2006). 
The sample of clusters with the calculated orbital elements turns 
out to be approximately a factor of 4 smaller in size, but, 
nevertheless, it is deemed representative relative to all of the 
observed Galactic clusters (Wu et\,al.~2009). Figure 1a shows the 
central part of the ''Galactocentric distance projected onto the 
Galactic plane - distance from the Galactic plane'' diagram for all of 
the optically visible clusters (Dias et\,al.~2002), while the open 
circles highlight the clusters of our catalogue. The brightest feature 
for both catalogues is the semicircular region of an enhanced density 
of points. We clearly see that within 
$\pm 1$~ kpc (designated by the two vertical dotted lines) from the 
solar orbital radius, a high density of points extends up to 
$|z| \approx 0.15$~kpc, while this density begins to rapidly drop 
outside this range. The effect is attributable to an enhanced 
interstellar reddening in the Galactic plane, which makes it difficult 
to determine the distances to clusters lying near this plane. It can 
also be seen from the diagram that the clusters in our catalogue are 
observed at progressively larger distances from the Galactic plane with 
increasing Galactocentric distance (the most distant clusters are 
outside the diagram). The lower density of points in the upper left 
corner of the diagram is not associated with observational selection, 
because the reddening is low at high Galactic latitudes even in the 
inner disk regions and, therefore, the distances to high clusters are 
determined fairly reliably. This suggests that although the selection 
effects make it difficult to reveal clusters lying near the Galactic 
plane, they, nevertheless, do not prevent the detection of very distant 
clusters. Therefore, we have the right to use our catalogue to analyze 
the properties of open clusters with different spatial-kinematical 
parameters. In particular, it can be seen from the figure that the 
maximum values of $|z|$ also increase with Galactocentric distance. 
This is because the gravitational potential in the Galactic plane 
decreases in this case.

{\bf\em Galactic Orbital Elements.} 

It is generally believed that the open star clusters are born from 
the interstellar matter distributed in the form of a thin layer in 
the Galactic plane. The scale height of this layer in the solar 
neighborhoods is variously estimated to be within the range 
50 - 75~pc (see, e.g., Malhotra 1994; Vergely et\,al. 1998; 
Weiss et\,al.~1999). Since this matter moves around the Galactic 
center almost in circular orbits, the open clusters should be 
expected to follow them as well. However, several clusters are in 
highly eccentric orbits that often rise high above the Galactic plane, 
which is indicative of their ''unusual'' origin 
(Vande Putte et\,al.~2010). In order to separate the clusters of 
different natures from one another, let us assume that only the 
clusters with circular low orbits were formed from the interstellar 
matter of the Galactic thin disk, for example, under the action of 
spiral density waves producing a shock wave traveling parallel to the 
Galactic plane, while the clusters with eccentric high orbits were 
formed through the action of other mechanisms on this matter. The idea 
that all clusters are formed in identical processes from an 
interstellar medium whose turbulence decreases with time serves as an 
alternative to this assumption. However, the latter assumption is in 
conflict with the theoretical calculations suggesting that the time of 
free collapse of a dissipating rotating protogalactic cloud into a flat 
disk is less than 0.4 Gyr; therefore, the clusters with the highest 
eccentric orbits must be the oldest. In fact, as can be seen from the 
table, the clusters with $Z_{max} > 8$ kpc have ages less than 2~Gyr, 
i.e., they are several times less than the age of the thin--disk 
subsystem. In addition, the velocity dispersions in clusters turn out 
to be considerably higher than those in field thin--disk stars 
(for more details, see below). Conceptually, this must suggest greater 
turbulence of the interstellar medium at the instant the clusters were 
formed from it and, hence, their older age. These inconsistencies force 
us to reject the alternative in favor of the originally advanced 
hypothesis suggesting the existence of open clusters of an unusual origin.

The thin--disk stars are known to be characterized by low residual 
velocities relative to the LSR and by almost circular orbits all points 
of which do not rise high above the Galactic plane. Nearby stars can be 
stratified with confidence into Galactic subsystems by their space 
velocity components relative to the LSR 
(see, e.g., Koval' et\,al.~2009). However, for more distant objects 
whose heliocentric distances are comparable to the solar Galactocentric 
distance, it is more reliable to determine their membership in a 
subsystem from Galactic orbital elements. The eccentricity (e) and 
the maximum distance of the orbital points from the Galactic plane 
$(Z_{max})$ (see, e.g., Marsakov and Borkova 2005; 
Vande Putte et\,al.~2010) are the most informative (in this respect) 
orbital elements. The indicator composed of the same orbital elements 
and proposed by Chiappini~et\,al.~(1997), $(Z^2_{max}+4e^2)^{1/2}$, 
where $Z_{max}$ is measured in kpc, turned out to be very convenient. 
Figure 1b presents the $Z_{max}-e$ diagram for all clusters of our 
catalog. The almost circular region with a higher density in the lower 
left corner of the diagram engages our attention. The clusters in this 
region can be separated by the boundary value of 
$(Z^2_{max}+4e^2)^{1/2} = 0.35$ (see the solid curve in the diagram): 
the density of points in the diagram outside this radius is an order of 
magnitude lower than that in the central region of the concentration. 
Most ($>80\,\%$) of the open clusters are inside the curve, implying 
that they reflect the kinematical properties typical of Galactic--disk objects.
Figure 1c shows the distribution of the selected clusters in their 
current distance from the Galactic plane taken in absolute value. 
(Recall that Wu et al. (2009) analyzed the incompleteness of the sample 
of clusters with known orbital elements and reached the conclusion 
about the legitimacy of using it to estimate the characteristic 
parameters of this population.) The derived distribution was fitted by 
an exponential law: $n(z) = Ce^{-Z/Z_o}$ ,where $Z_0$ is the scale 
height. The scale height for this subsample of clusters turned out to 
be approximately the same as that for the interstellar medium in the 
solar neighborhoods, $Z_0 = (65 \pm 4)$~pc. (Within 1~kpc, where the 
sample may be considered essentially complete, the clusters with such 
circular low orbits show $Z_0 = (70 \pm 12)$~pc, i.e., the same value, 
within the error limits.) The selection conditions suggest that the 
clusters of this subgroup were formed from a kinematically cold 
interstellar medium. However, all stars of the Gould Belt and the 
comparatively young Hyades--Pleiades star stream, which is believed to 
have resulted from the perturbation of the interstellar medium by 
spiral density waves, satisfy these constraints. In other words, the 
criterion (Z2max + 4e2)1/2 < 0.35 turns out to correspond well to the 
young stellar population of the thin disk, and we believe the clusters 
satisfying it to be typical of the Galactic disk. Therefore, it seems 
natural to assume that all clusters of this kinematically cold group 
were formed from the matter reprocessed exclusively in genetically 
related stars of the Galaxy, i.e., from the matter of a single 
protogalactic cloud. However, it follows from Figs. 2a and 2b that this is
not quite the case. These figures show the ''$Z_{max}$~-- [Fe/H]'' and
''e~-- [Fe/H]'' diagrams, respectively. The filled circles highlight 
the ''kinematically cold'' clusters satisfying the criterion 
$(Z^2_{max}+4e^2)^{1/2} = 0.35$. We see that such clusters fill a wide 
metallicity range. Thus, it turns out that there are such clusters 
whose low metallicities are atypical of the local field thin-disk stars 
even among the clusters with flat circular orbits. The clusters with 
eccentric high orbits show a tendency for their metallicity to decrease 
starting from the solar one as both orbital elements increase 
(in Fig.~2a, three clusters with $Z_{max} > 8$~kpc are outside the 
diagram and are disregarded when constructing the regression line). 
Such dependences may suggest that an increasing fraction of 
interstellar matter with a low metallicity is involved in the 
formation of clusters with increasingly eccentric high orbits. The 
general appearance of Figs.~2a and~2b differs from the analogous 
diagrams from Vande Putte et\,al.~(2010), because low--metallicity 
($[Fe/H] < -0.4$) clusters with circular low orbits are absent in the 
latter paper. The reason is that all these clusters are distant and lie 
essentially in the Galactic plane, where the interstellar extinction, 
which makes it difficult to determine the photometric metallicity used 
by these authors, is great. We took the data for these clusters from 
Paunzen (2010) and version 3.1 of the catalogue by Dias et\,al.~(2002) 
published already after the paper by Vande Putte et\,al.~(2010).

{\bf\em Metallicities.} 

Wu et\,al.~(2009) used the metallicity as an indicator for 
revealing clusters of an unusual origin. Let us examine the 
distribution of the open clusters in heavy-element abundances in 
our catalogue. Figure 2c shows the metallicity function for 264 
clusters. The solid curve indicates the fit to the histogram by 
the sum of two Gaussians. The Gaussian parameters determined by 
the maximum likelihood method showed that the probability of 
erroneously rejecting the hypothesis about the description of the 
distribution by one Gaussian against the alternative of its 
representation by the sum of two Gaussians is $<5\,\%$ 
(for the method, see Martin 1971). As a result, we see that 
the population of open clusters turns out to be heterogeneous 
in metallicity as well, and it can be divided into two groups 
by $\textrm{[Fe/H]} \approx -0.12$. The group with an approximately 
solar metallicity shows a low dispersion, while the low--metallicity 
group occupies a fairly wide [Fe/H] range. For comparison, Fig.~2c 
also plots the histogram for nearby field thin-disk stars with the 
photometric metallicities from Marsakov et\,al.~(2011). For 
convenience, both histograms were normalized to the total numbers of 
corresponding objects. We see that the metallicity functions for the 
clusters and field stars differ in general appearance. In particular, 
the metallicity range for the clusters is slightly larger toward 
negative values. Nevertheless, the principal maximum of the metallicity 
distribution for the clusters is more to the right and lies near 
$\textrm{[Fe/H]} = +0.05$, whereas for the field stars 
$\textrm{[Fe/H]} = -0.12$ (i.e., where the dip is observed for the 
clusters). However, as we see, there is a second, lower--metallicity 
maximum near $\textrm{[Fe/H]} = -0.27$ for the clusters. (Note that if 
[Fe/H] for the field stars and open clusters are reduced to the solar 
Galactocentric distance by correcting them for the radial metallicity 
gradient and by assuming the mean orbital radii to be their 
birthplaces, then the shape of the histograms being compared in 
Fig.~2c will not change fundamentally.)


\section{THE AGE DEPENDENCE OF THE SPATIAL-KINEMATICAL AND PHYSICAL PARAMETERS OF OPEN CLUSTERS}

{\bf\em Heliocentric Distances} 

Let us examine the extent to which evolutionary changes in clusters 
affect their detectability among the field stars. Figure~3a shows the 
''age--heliocentric distance'' diagram for all of the known open 
clusters. We see that the field is filled very nonuniformly and a dense 
cluster of points in the shape of a triangle is observed in the lower 
left corner of the diagram, while the upper right corner is almost 
empty. The existence of the upper inclined straight-line boundary 
separating the dense cluster of points in the lower left corner of the 
diagram from the remaining points is caused, on the one hand, by an 
enhanced brightness of young hot stars in the clusters: the younger the 
cluster, the greater the distance at which it can be distinguished 
against the background of Galactic field stars. On the other hand, the 
existence of this boundary is attributable to an intensive disruption 
of clusters reducing their total luminosity. Although young hot stars 
in clusters vanish with age, a sufficient number of red giants appear 
instead of them; therefore, the cluster luminosity increases again. 
As a result, we see the most distant clusters in the range 1 - 5 Gyr. 
(However, it should be noted that the z coordinates are generally also 
larger for distant clusters (see Fig.~1a), so that interstellar 
reddening does not hinder the distance determination for them.) At the 
same time, however, an intense loss of cluster stars through dynamical 
processes begins to have an effect at an even older age, and, in the 
long run, they all dissipate completely. The dissipation process 
continues over the entire cluster lifetime, causing their number in the 
Galaxy to decrease exponentially with time.

{\bf\em Residual Velocity Components.} 

The velocity dispersions of field Galactic-disk stars are 
known to increase with age. For all residual velocity components 
of the stars relative to the LSR, this dependence is well described 
by a power law of the form $\sigma_v \sim t^{0.25}$
(see, e.g., Koval' et\,al.~2009). The ''heating'' of the disk stars
by spiral density waves is believed to be most likely responsible 
for this increase. Let us estimate the extent to which the velocity 
dispersion of more massive objects, which the open clusters are, 
depends on the age. In contrast to nearby stars, to analyze the 
properties of such distant objects as open clusters, it is more 
appropriate to use the space velocity components in cylindrical 
coordinates. Figure~3b shows the ''age-azimuthal velocity'' diagram 
for the clusters of our catalogue. (To save space in the diagram, 
the $V_R$ and $V_Z$ velocity components are not presented, but 
morphologically they have an appearance very similar to Fig.~3b.) 
The region distinguished by an enhanced density of points is in the 
shape of a triangle (see the dashed lines drawn by eye) with the 
vertex lying on the line of the mean azimuthal velocity for these 
clusters and with an abscissa of $\approx 1$~ Gyr. Just as in 
Fig.~3a, the existence of the inclined density boundaries in the 
diagram is attributable the decrease in cluster luminosity with age. 
Note that, in this case, the total scatter of velocities is essentially 
independent of the age, suggesting that the cluster space velocities 
(as in the case of field stars) do not increase with time. However, the 
velocity dispersions actually increase rapidly with age, reaching 
$(\sigma_\Pi,\sigma_\Theta,\sigma_Z)= (49, 115, 58)$~~km\,s$^{-1}$ 
for the oldest clusters (see the dashed 
lines in Fig. 3c), whereas the limiting dispersions, on average, for 
considerably older nearby field thin-disk stars are significantly 
lower $\sigma_U, \sigma_V, \sigma_W) = (44, 27, 21)$~km\,s$^{-1}$  
(see Koval' et\,al.~2009). To construct Fig. 3c, we divided all 
clusters of our catalogue into seven age subgroups decreasing in number 
(from 155 for the youngest ones to 13 for the oldest ones). The number 
decreased monotonically in order that it could be possible to clearly 
trace the behavior of the cluster velocity component dispersions in 
the entire range of ages occupied by them nonuniformly. It is generally 
believed that the increase in velocity dispersion with age for clusters 
is caused either by relaxation effects (see, e.g., Wu et\,al.~2009) or 
by a decreasing (with time) degree of turbulence of the interstellar 
medium in the thin disk. We see from Fig.~3b that the progressive 
decrease in the density of points near the average line in the diagram, 
which roughly corresponds to the circular velocity of the Galaxy at the 
solar Galactocentric distance, causes the dispersion to increase. This 
is most likely because clusters with nearly circular orbits are 
intensively disrupted as a result of their long stay near massive 
clouds of interstellar matter (see, e.g., Danilov and Seleznev 1994) 
and spiral density waves (Mishurov and Acharova 2011) concentrating 
near the Galactic plane.

Some decrease in dispersion within the first several hundred Myr, 
which is especially noticeable for the $V_\Theta$ velocity component, also
engages our attention in Fig.~3c. This effect is most likely caused by 
the observational selection effect associated with the differential 
rotation of the Galactic disk and the high luminosity of the young 
clusters visible at significantly differing Galactocentric distances. 
To test this assumption, Fig.~3c presents the age dependences of the 
velocity dispersions for the clusters lying within 1 kpc of the Sun. 
Recall that the original sample of clusters within this range may be 
considered essentially complete (see Piskunov et\,al.~2006). 
(We divided this sample into six age ranges; in the youngest group, 
there were several times more stars than in the oldest one, where there 
are only 11 of them.) As we see, the dispersions of all velocity 
components for nearby clusters a $t<1$~Gyr actually do not depend on 
the age, within the uncertainty limits. Note that the low dispersions 
in the oldest group were obtained, because it contains no very old 
clusters (see Fig.~3a) and clusters lying high above the Galactic 
plane, and precisely they provide high velocity dispersions.

{\bf\em Physical Parameters} 

In Fig.~4a, the masses of open clusters are plotted against their 
age. (The three oldest clusters are not shown in the diagram.) We 
see that this plot slightly differs in appearance from the plots of 
spatial-kinematical parameters against the age. Here, the scatter is 
largest for the youngest clusters and decreases with age, which allows 
the inclined upper boundary to be drawn. The location of the lower 
boundary is virtually independent of the age. As a result, it turns out 
that the most massive clusters are among the youngest ones, and the 
upper cluster mass boundary sinks with increasing age. The dots inside 
circles designate the clusters lying within 1~kpc of the Sun. We see 
that they fill rather uniformly the entire field in the diagram, except 
for the upper left corner, where the most massive, youngest, and, as we 
see, most distant clusters are found. This suggests that not the 
distance selection causes the inclination of the upper envelope 
observed in the diagram, but it is entirely attributable to 
evolutionary changes in clusters. The central concentration and 
ellipticity distributions in Figs.~4b and~4c exhibit no age dependences.

In Fig.~4d, the dispersions of the physical clusters parameters are 
plotted against the age. We see that the mass dispersion decreases 
within the first $\sim 200$~Myr and then increases again. Such a 
behavior probably stems from the fact that one of the two mechanisms 
prevails for clusters of different ages. On the one hand, from their 
very birth, the open clusters begin to lose their stars; that is why 
the upper limit on their mass decreases. On the other hand, as the age 
increases, the low-mass clusters begin to be disrupted completely with 
acceleration; therefore, the mean mass of the survived clusters turns 
out to be higher. As we see, the ellipticity and central concentration 
dispersions remain within the range of age--independent uncertainties. 
Thus, evolutionary changes are detected only for the total integrated 
mass of the clusters.

In the next paper (Gozha et\,al.~2012), we will formulate the 
principles of identification of open clusters of different natures and 
investigate the properties of both populations.

\section*{ACKNOWLEDGMENTS}

We are grateful to V.M.~Danilov and A.V.~Loktin for a preliminary 
familiarity with the manuscript and constructive additions as well 
as to A.E. Piskunov for useful consultations. This work was supported 
by the Russian Foundation for Basic Research (project no. 11-02-00621 
a). V.A.~Marsakov also thanks the Ministry of Education and Science of 
the Russian Federation for support (project P 685).

\renewcommand{\refname}{Список литературы}

\newpage

\begin{figure*}
\centering
\includegraphics[angle=0,width=0.96\textwidth,clip]{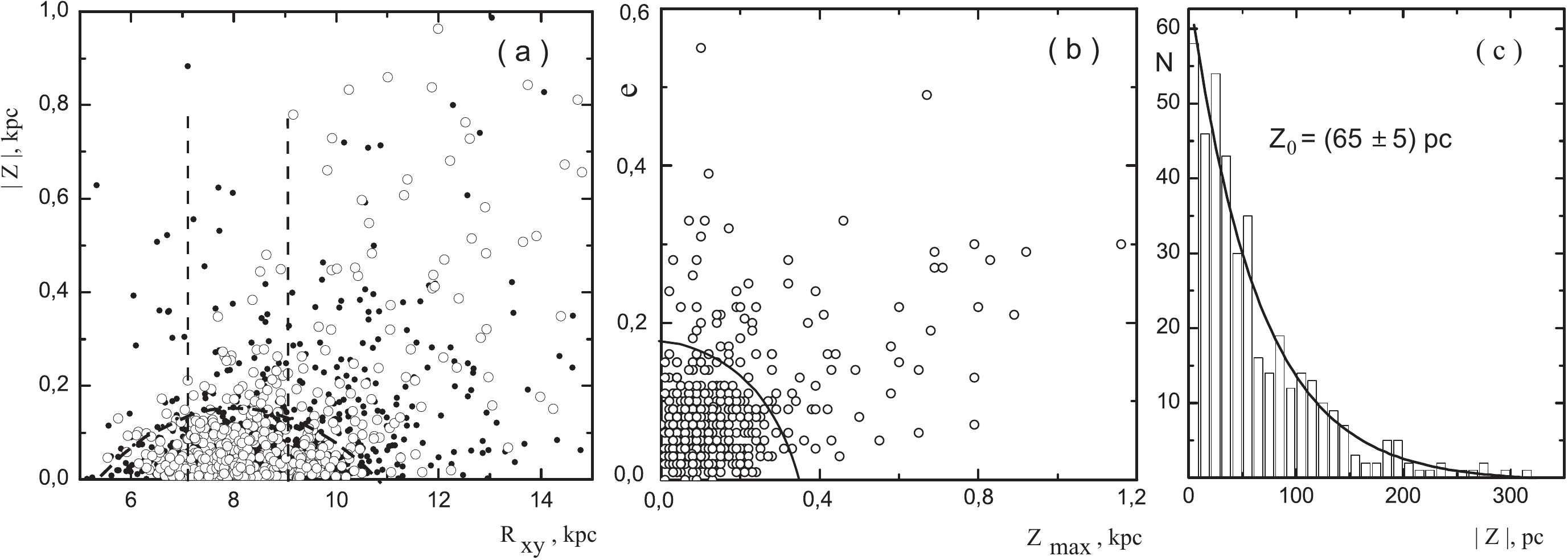}
\caption{(a) ''Position projected onto the 
     Galactic plane - height above the Galactic plane'' 
     diagram for open clusters: the filled and open circles 
     represent the clusters from the catalogue by 
     Dias~et\,al.~(2002) and our catalogue, respectively. 
     (b) ''Maximum distance of the orbital points from the 
     Galactic plane - orbital eccentricity'' diagram for the 
     clusters of our catalogue; the part of the circumference 
     with the radius $(Z^2_{max} + 4e^2)^{1/2} = 0.35$ separates 
     out the region of an enhanced density of points in the diagram. 
     (c) The distribution of clusters satisfying the criterion 
     $(Z^2_{max} + 4e^2)^{1/2} < 0.35$ in current distance from the 
     Galactic plane taken in absolute value; the solid curve indicates 
     an exponential fit to the distribution and the scale height with 
     its uncertainty is indicated.}
\label{fig1}
\end{figure*}

\newpage

\begin{figure*}
\centering
\includegraphics[angle=0,width=0.90\textwidth,clip]{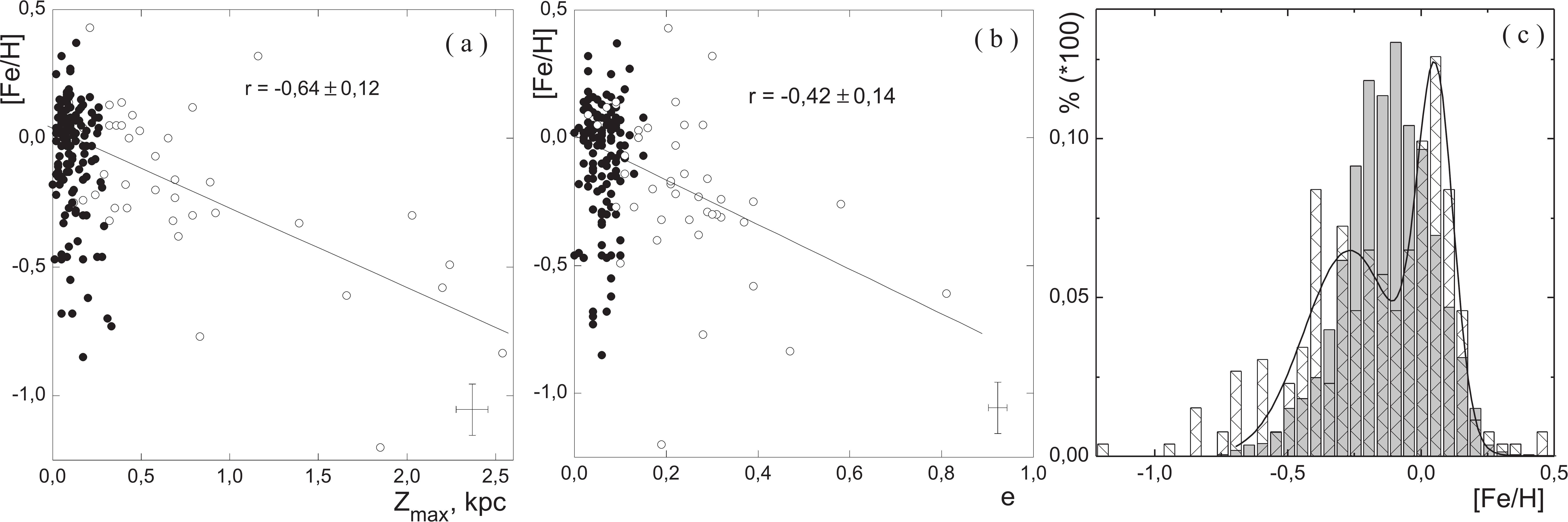}
\caption{Cluster metallicity versus maximum distance of 
     the orbital points from the Galactic plane (a) and versus 
     orbital eccentricity (b); (c) the metallicity distributions 
     of open clusters (hatched) and field thin-disk stars (gray color). 
     The filled and open circles represent the clusters satisfying the 
     criterion $(Z^2_{max} + 4e^2)^{1/2} < 0.35$ and the remaining 
     clusters of the catalogue, respectively; the lines represent the 
     regression lines for the open circles; the correlation 
     coefficients are indicated. The curve in panel (c) is the fit 
     to the [Fe/H] distribution of clusters by the sum of two Gaussians.}
\label{fig2}
\end{figure*}

\newpage

\begin{figure*}
\centering
\includegraphics[angle=0,width=0.96\textwidth,clip]{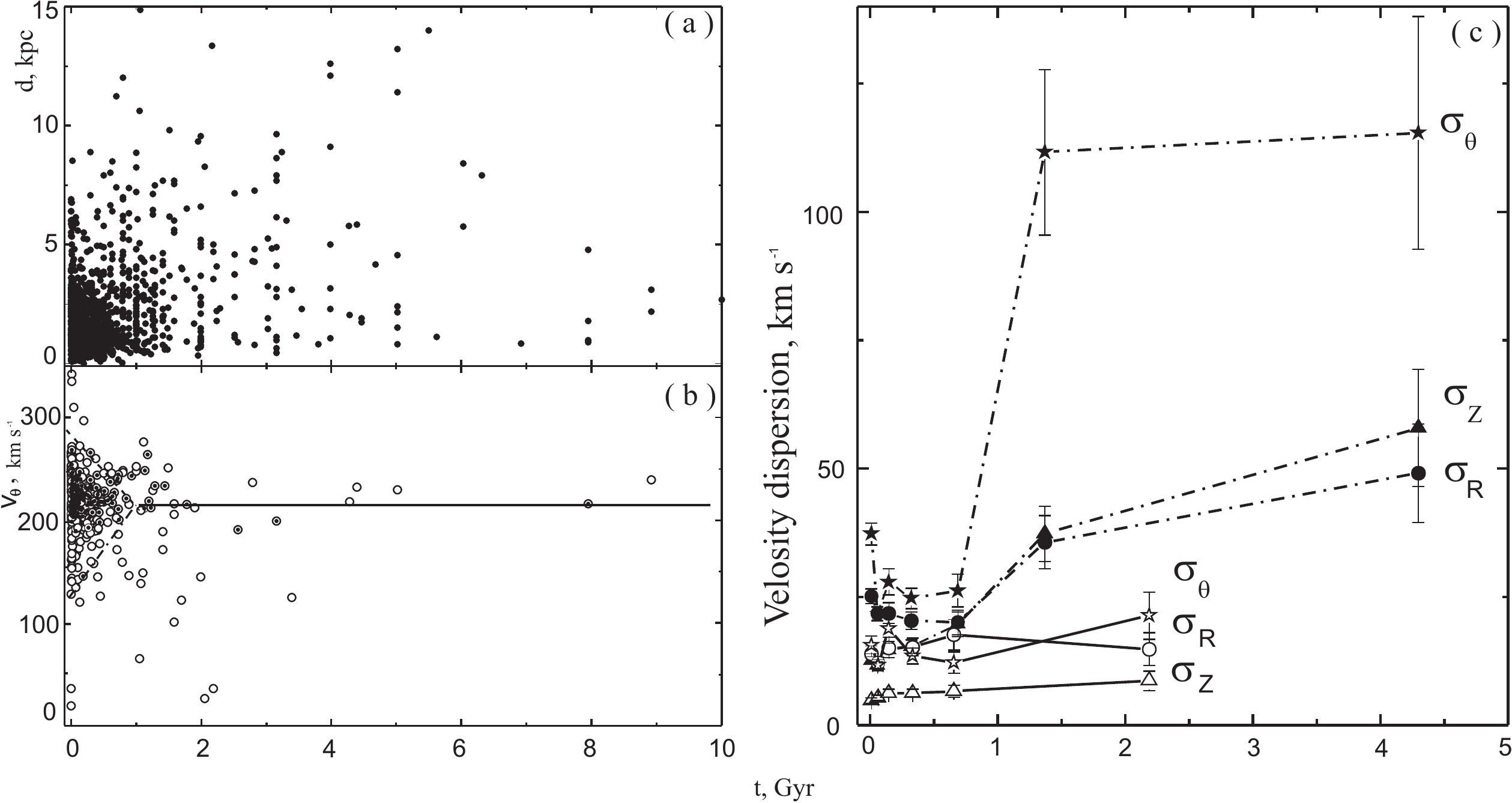}
\caption{Heliocentric distances for all optically visible 
     open clusters (a), azimuthal space velocity components (b), 
     and dispersions $\sigma_\Pi,\sigma_\Theta,\sigma_Z$ of the 
     cluster velocity components (c) versus age. In panel (b), 
     the open circles and the dots 
     inside circles are the clusters of our catalog and those lying 
     within 1~kpc of the Sun; the horizontal dashed line indicates 
     the mean rotation velocity of the clusters around the Galactic 
     center; the inclined dashed lines are the straight lines 
     separating the region of an enhanced density of points in the 
     diagram that were drawn by eye. In panel (c), the dashed and solid 
     polygonal lines indicate the dispersion variations for the 
     clusters of our catalogue and those lying within 1~kpc of the Sun, 
     respectively; the errors of the dispersions are shown.}
\label{fig3}
\end{figure*}

\newpage

\begin{figure*}
\centering
\includegraphics[angle=0,width=0.96\textwidth,clip]{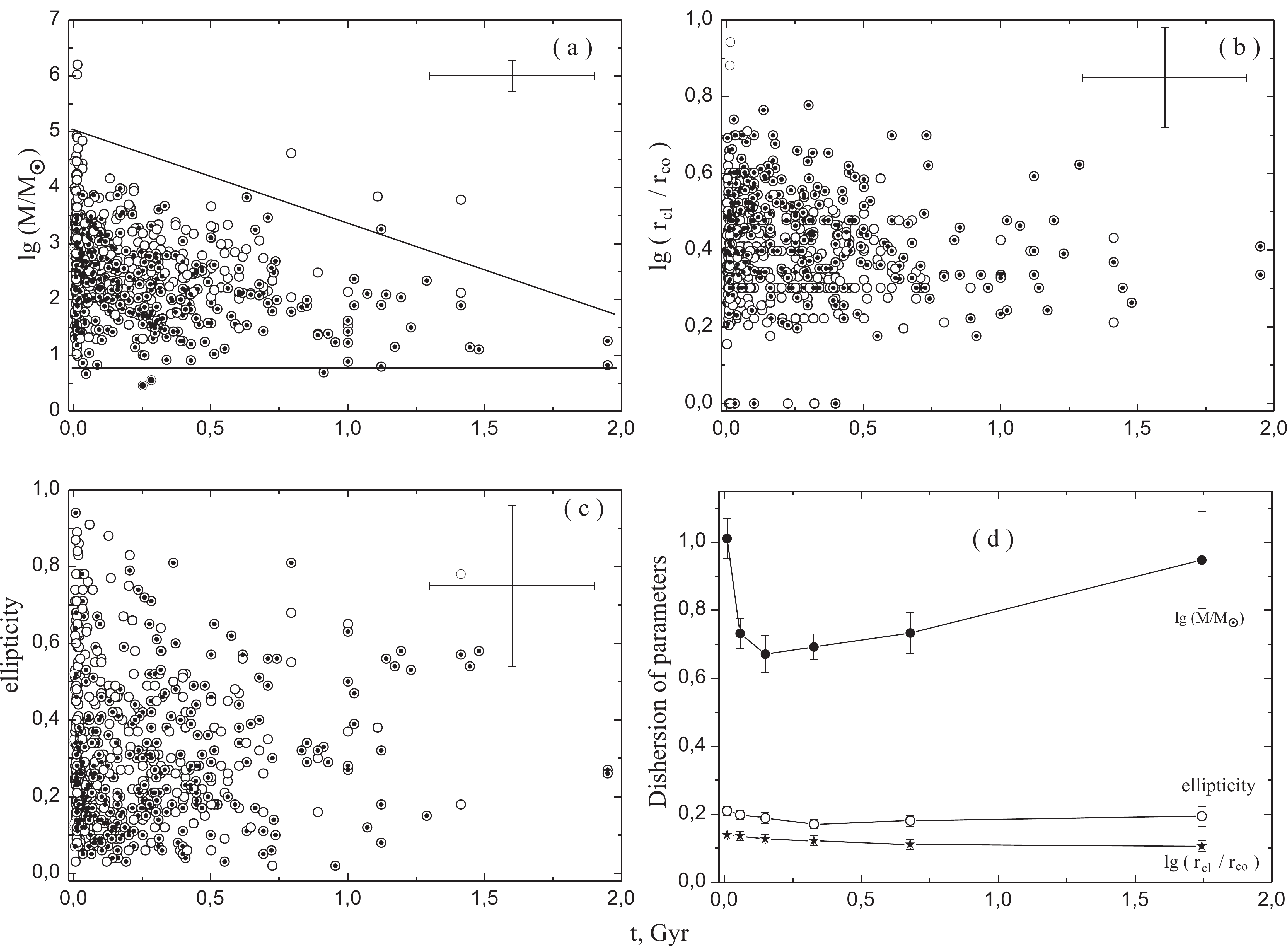}
\caption{Masses (a), central concentrations (b), 
     ellipticities (b), and dispersions of these three 
     parameters (d) versus age for open clusters. The dots 
     inside circles highlight the clusters lying within 1~kpc 
     of the Sun. In panel (a), the dashed lines indicate the 
     upper and lower envelopes drawn by eye.}
\label{fig4}
\end{figure*}

\end{document}